\documentclass[12pt]{article}
\usepackage{psfig}
\def\lesssim{\mathrel{\mathpalette\vereq<}}
\def\gtrsim{\mathrel{\mathpalette\vereq>}}
\def\lsim{\raise0.3ex\hbox{$\;<$\kern-0.75em\raise-1.1ex\hbox{$\sim\;$}}}
\def\gsim{\raise0.3ex\hbox{$\;>$\kern-0.75em\raise-1.1ex\hbox{$\sim\;$}}} 
\def\21{$SU(2)_L \otimes U(1) $}
\def\ne{\hbox{$\nu_e$ }}
\def\nm{\hbox{$\nu_\mu$ }}
  
\makeatletter
\def\vereq#1#2{\lower3pt\vbox{\baselineskip1.5pt \lineskip1.5pt
\ialign{$\m@th#1\hfill##\hfil$\crcr#2\crcr\sim\crcr}}}
\makeatother

\begin{document}
\begin{titlepage}
\begin{center}
\hfill    CERN-TH/2000-319\\
\hfill    IFIC/00-63\\
~{} \hfill hep-ph/0010299\\

\vskip .3in
{\large \bf Minimalistic Neutrino Mass Model}

\vskip 0.5in

Andr\'e de Gouv\^ea

\vskip 0.1in

{\em CERN - Theory Division\\
     CH-1211 Gen\`eve 23, Switzerland}

\vskip .2in

Jos\'e W. F. Valle

\vskip 0.1in

{\em    Instituto de F\'{\i}sica Corpuscular~--~C.S.I.C., 
Universitat de Val\`encia, \\
Edificio Institutos de Paterna, Apt.~22085, E-46071 Val\`encia, Spain}

\end{center}

\vskip .2in

\begin{abstract}
  We consider the simplest model which solves the solar and
  atmospheric neutrino puzzles, in the sense that it contains the
  smallest amount of beyond the Standard Model ingredients. The solar
  neutrino data is accounted for by Planck-mass effects while the
  atmospheric neutrino anomaly is due to the existence of a single
  right-handed neutrino at an intermediate mass scale between
  $10^9$~GeV and $10^{14}$~GeV.  Even though the neutrino mixing
  angles are not exactly predicted, they can be naturally large, which
  agrees well with the current experimental situation.  Furthermore,
  the amount of lepton asymmetry produced in the early universe by the
  decay of the right-handed neutrino is very predictive and may be
  enough to explain the current baryon-to-photon ratio if the
  right-handed neutrinos are produced out of thermal equilibrium.  One
  definitive test for the model is the search for anomalous seasonal
  effects at Borexino.
\end{abstract}

\end{titlepage}

\newpage
\setcounter{footnote}{0}
\setcounter{equation}{0}
\section{Neutrino Puzzles and  Masses}

The SuperKamiokande atmospheric neutrino data \cite{SuperK-atmos},
provide very strong evidence for neutrino conversion, {\it i.e.}\/ that
neutrinos produced in well defined flavour eigenstates partially
convert to different flavour eigenstates at the detection point.
A similar conclusion follows from the long standing solar neutrino
puzzle \cite{solar} and a number of other atmospheric neutrino
experiments \cite{atmos}.
Altogether, solar and atmospheric data indicate the need for \ne and
\nm conversions, respectively.
There are a number of neutrino flavour conversion
mechanisms~\cite{solrsf,solexotic,atmexotic}, and the simplest one is
that of neutrino oscillations~\cite{oscreview}. 

Although in some circumstances neutrino oscillations do not imply
neutrino masses~\cite{Valle:1987gv}, typically they do. Due to the
fact that neutrino mass and flavour eigenstates should, in general,
differ, quantum interference effects can lead to sizeable
neutrino conversion rates either from large mixing angles or from
resonant matter effects. In this scenario, the solar and atmospheric
neutrino puzzles are interpreted as strong evidence for nonzero
neutrino masses.

The present data can be summarised as follows:
the atmospheric neutrino puzzle is best solved by $\nu_{\mu} \to
\nu_{\tau}$ with close to maximal mixing ($\sin^2\theta \simeq 0.5$)
and $\Delta m^2_{\rm atm} \simeq
10^{-3}-10^{-2}$~(eV)$^2$~\cite{latestglobalanalysis}.  On the other
hand the solar neutrino puzzle requires $\nu_e \to \nu_x$
oscillations, where $\nu_x$ is some linear combination of $\nu_{\mu}$
and $\nu_{\tau}$.
While the atmospheric neutrino data indicate large mixing angles, the 
current solar neutrino data do not indicate a unique solution.
The mixing is either very small ($\sin^2 \omega \simeq 10^{-3}$) or
quite large ($0.1 \lesssim \sin^2 \omega \lesssim 0.5$), while the
value of $\Delta m^2_{\odot}$ varies from $10^{-4}$~eV$^2$ to
$10^{-10}$~eV$^2$~\cite{latestglobalanalysis}.

The neutrino oscillation interpretation of the solar and atmospheric data
only fixes the mass-squared splittings amongst the the neutrinos, not the
overall mass scale. In fact, the data are equally well explained if all
neutrino masses were almost degenerate at some arbitrarily large value.
There are, however, constraints on neutrino masses from tritium
beta decay experiments at the level of 3 eV~\cite{PDG}, while
cosmological considerations impose limits on the sum of all stable
neutrino species, namely $\sum_{i} m_{\nu}^i\lesssim 30$~eV. There are
also a number of results from the non-observation of neutrinoless
double beta decay \cite{double-beta}, which imply that $M_{ee}=\sum_i
U_{ei}^2 m_{\nu_i} \lsim 0.2$ eV where $\nu_i$, $i=1,2,3$ are the
neutrino mass eigenstates, with masses $m_{\nu_i}$, and $U_{ei}$ are
the components of $\nu_e$ in the $\nu_i$ basis ($\nu_e=\sum_i U_{ei}
\nu_i$).

Barring the possibility of quasi-degenerate neutrinos an/or exotic
neutrino conversion mechanisms, all existing data\footnote{We will
  neglect throughout the LSND \cite{LSND} result which, if confirmed
  by an independent experiment, would lead to the existence of light
  sterile neutrinos~\cite{4nu,4nubrane}. We therefore only consider
  the possibility of 3 light neutrino species.}
are easily satisfied if one assumes that the heaviest neutrino weighs
around $0.05$~eV and is an almost fifty-fifty mixture of $\nu_{\mu}$
and $\nu_{\tau}$, while the second to heaviest neutrino weighs from
$10^{-2}$~eV to $10^{-5}$~eV and is either composed of mostly the
orthogonal combination of $\nu_{\mu}$ and $\nu_{\tau}$ with a very
small $\nu_e$ component or an almost fifty-fifty mixture of this
$\nu_{\mu}$ and $\nu_{\tau}$ linear combination and $\nu_e$.  There is
no preferred value for the mass of the lightest mass eigenstate, which
can even be massless as far as the experimental constraints are
concerned. In this case neutrino masses are very small, with the
heaviest neutrino about 10$^{7}$ times lighter than the lightest
charged fermion.

What is wrong with neutrino masses? Na\"{\i}vely, they are not allowed
by the Standard Model (SM). This is very easy to see: given the field
content and the gauge symmetries of the SM, it is impossible to write
down a renormalizable neutrino mass term.  However, if the SM is
regarded as the effective low-energy limit of some, unknown, higher
energy theory, neutrino Majorana masses are not forbidden and do arise
after electroweak symmetry breaking. The neutrino Majorana mass comes
from a dimension-5 term in the Lagrangian\footnote{An alternative
  approach generates Majorana neutrino masses by extending the SM
  Higgs sector to include exotic representations such as $SU(2)_L$
  triplets with small vacuum expectation
  values~\cite{Schechter:1980gr}. We will not consider this approach
  here.}
\begin{equation}
{\cal L}_5= \frac{\lambda^{ij}}{M}(L^iH)(L^jH),
\label{5-dim}
\end{equation} 
such that, after electroweak symmetry breaking, one generates a
neutrino mass matrix, $m_{\nu}^{ij}$ given by
\begin{equation}
m_{\nu}^{ij}=\frac{\lambda^{ij}v^2}{M}.
\label{mass}
\end{equation}
Here $L^i$ is a lepton doublet field ($i,j=e,\mu,\tau$), $H$ is the
Higgs doublet field and $M$ is an arbitrary mass scale. 
It is somehow related to
the unknown high energy theory of which the SM is an effective theory.
The $\lambda^{ij}$ are dimensionless couplings and $v=174$~GeV is the
Higgs vacuum expectation value (vev). Note that Eq.~(\ref{5-dim})
leads to a Majorana mass term which explicitly violates the lepton
number symmetry, $L$.  This is not a surprise since both lepton and
baryon number ($B$) are accidental global symmetries of the SM, and
there is no reason they should be conserved by the higher energy
theory. As a matter of fact, $B+L$ is violated in the SM via
electroweak quantum effects, while $B-L$ is still conserved.
Furthermore, if string theory has anything to do with reality, it
seems to predict that there are no exactly conserved global
symmetries, and the only way to realize $B-L$ as a conserved symmetry
is to assume it is gauged.

Alternatively, renormalizable neutrino Dirac masses are easily allowed
within the SM if right-handed fermions uncharged under the SM gauge
group exist.  These are often referred to as right-handed neutrinos,
and, even though they are not present in the SM, they can be added
without causing any trouble (as they are SM singlets). Furthermore,
their existence is quite natural in the context of a number of
extended electroweak models as well as certain grand unified theories
(GUTs)~\cite{threview}. In their presence, the neutrinos acquire a
Dirac mass like all the charged fermions, from a Yukawa coupling to
the Higgs field
\begin{equation}
{\cal L}_{LNH}=y^{ik} L^{i} N^{k} H + h.~c.~,
\label{yukawa}
\end{equation}
where $N^{k}$ ($k=1...n$, where $n$ is the number of right-handed
neutrinos) are right-handed fermions.  As far as the SM gauge group is
concerned, $n$ is completely unconstrained, since $N^k$ are \21
singlets~\cite{Schechter:1980gr}.  The $y^{ik}$ are the couplings of
the sterile neutrinos to the lepton and Higgs doublet fields, leading
to a neutrino Dirac mass matrix given by $y^{ik}v$.

A crucial fact is that right-handed neutrinos can also have Majorana
masses, which are unprotected by the gauge symmetry. This means that
the right-handed neutrino Majorana mass term comes from a
renormalizable term in the Lagrangian (indeed from a relevant
operator), which is completely unrelated to the electroweak symmetry
breaking scale. The terms involving the right-handed neutrino are
\begin{equation}
V_{N} = y^{ik} L^{i} N^{k} H + \frac{1}{2}M_N^{kl} N^k N^l + h.~c.~,
\label{right-handed}
\end{equation} 
where $M_N^{kl}$ is the right-handed neutrino Majorana mass matrix.
The simultaneous presence of $y^{ik}$ and $M_N^{kl}$ in
Eq.~(\ref{right-handed}) implies that $L$ is violated.

In summary, the presence of right-handed neutrinos not only allows a
Dirac mass term for the neutrino sector, but, in general, also
violates $L$. Most importantly, it introduces a new energy scale into
the SM Lagrangian. The task of neutrino mass modeling is then to
explain the origin of Eq.~(\ref{5-dim}) and of the $\lambda^{ij}$ and
$M$ values or, alternatively, to determine the $y^{ik}$ and $M_N$
values in Eq.~(\ref{right-handed}).

In this paper, we present a neutrino mass model which combines the
above alternatives to generate neutrino masses and contains the least amount
of new ingredients added to the SM. In Sec.~2 we define what is meant
by ``new ingredients'' (or physics beyond the SM), and argue in which
sense the solar and atmospheric neutrino puzzles imply new physics. In
Sec.~3 we present this ``minimalistic'' model and show how it
satisfies the current experimental data. In Sec.~4 we discuss whether
the minimalistic model allows for an explanation of the
baryon-to-entropy ratio of the universe, via the elegant and
economical mechanism of leptogenesis, and in Sec.~5 we conclude.

\setcounter{footnote}{0}
\setcounter{equation}{0}

\section{``How Much'' New Physics?}

Generating neutrino masses requires the addition of extra degrees of
freedom and/or interactions to the SM~\cite{threview}, {\it i.e.,}\/ new physics.  
New physics is
characterized by an energy scale above which the SM (and, perhaps,
quantum field theory) is no longer appropriate and a different theory
is required to describe physical processes involving this energy
scale. This, of course, is what is meant by saying that the SM is the
effective theory of some higher energy theory.

New physics usually manifests itself in the SM Lagrangian in the form
of irrelevant operators, suppressed by powers of the scale of new
physics.  The presence of these operators can be detected either via
the observation of rare or forbidden processes which are
suppressed in the SM thanks to accidental symmetries which need not be
preserved by the high energy theory, such as proton decay and $\mu \to
e\gamma$, or via precision measurements of SM observables.  

Independent of any hint for new physics, it is clear that the SM is an
effective theory because it disregards the gravitational force, which
certainly exists. This means that, {\sl if nothing else happens,}\/
the SM has to break down at the energy scale where gravitational
effects become relevant, {\it i.e.,}\/ the Planck scale, $M_{Pl}\simeq
10^{18}$~GeV. This means that, independent of whether there is some
``intermediate'' new physics, higher dimensional operators suppressed
by the Planck scale should be present, unless forbidden by some deep,
fundamental reason.

These ``$1/M_{Pl}$'' operators are harmless. For example, a
dimension-6 term $QQQL/M_{Pl}^2$ violates $B$ and $L$, and leads to
proton decay.  However, the expected proton lifetime due to this
operator is many orders of magnitude away from the current
experimental reach \cite{pdecay}. The same is true of operators which
mediate $\mu \to e\gamma$ and other rare/forbidden processes.

As first noted by Barbieri, Ellis, and Gaillard \cite{BEG} in the context of
$SU(5)$ GUTS, and further explored by the authors of \cite{ABS} in the case 
of the SM, ``quantum gravity'' effects also generate Eq.~(\ref{5-dim}), 
namely,
\begin{equation}
{\cal L}_{QG}= \frac{\lambda^{ij}}{M_{Pl}}(L^iH)(L^jH),
\label{1/M_Pl}
\end{equation} 
should be considered as a ``standard'' term of the SM Lagrangian. What
is most remarkable is that, for $\lambda^{ij}$ of order 1, the
entries of the Majorana neutrino mass matrix are of the order
$\frac{(100)^2} {10^{18}}\times 10^{9}$~eV~$=10^{-5}$~eV. This is in
the acceptable mass range which solves the solar neutrino
puzzle~\cite{latestglobalanalysis}! 
This implies the possibility that the solar neutrino puzzle can be
solved without assuming any new physics between the electroweak and
the Planck scale. In contrast, the atmospheric neutrino puzzle, when
interpreted as evidence for a neutrino mass, is unambiguous proof of
new physics, beyond the SM, at some intermediate energy scale.

An elegant way of understanding the smallness of neutrino masses is
the seesaw mechanism \cite{seesaw}, which takes advantage of
Eq.~(\ref{right-handed}). Since the right-handed neutrino mass scale
is expected to be much larger than the electroweak scale it can be
integrated out at energies smaller than this new mass scale, so that
the effective Lagrangian again looks like Eq.~(\ref{5-dim}), with
\begin{equation}
m_{\nu}^{ij}=y^{ik}(M_N^{-1})^{kl}(y^{T})^{lj}\times v^2.
\end{equation}
The order of magnitude for the elements of $M_N$ required to solve the
atmospheric neutrino puzzle is such that
$M_N=O(\frac{(yv)^2}{10^{-10}}{\rm GeV})$. For typical values of the Yukawa
couplings ($y$), one obtains $10^{9}$~GeV$\lesssim M_N\lesssim
10^{14}$~GeV.  

There are many other, indirect, hints for new physics.  Most
importantly, there is the problem of the stability of the electroweak
scale. Solving this so-called hierarchy problem, means, roughly,
understanding why the electroweak scale ({\it e.g.}\/ the SM Higgs
mass) is so much smaller than the Planck scale. Its most popular
solutions involve low-energy SUSY, which stabilizes scalar masses by
eliminating quadratic divergences or, more recently, the existence of
``large'' extra dimensions \cite{ADD}. The latter assumes that the
fundamental scale of quantum gravity is indeed the weak scale, and
that the apparent weakness of the gravitational force is due to the
fact that gravity propagates in a multidimensional space-time, while
SM fields are trapped in a four-dimensional space-time.

Other hints of new physics include the apparent unification of
gauge couplings at around $10^{16}$~GeV, which seems to be an
indication that the SM gauge group may be the ``left-over'' symmetry
of some high energy GUT symmetry, and the hierarchy of the charged
fermion masses which may be considered as a hint for horizontal flavour
symmetries broken at intermediate energy scales. 

In summary, there are many ``theoretical'' hints for new physics at very
different energy scales, in particular the weak scale ($M_W\simeq
10^{2-3}$~GeV) and the GUT scale ($M_{GUT}\simeq 10^{16}$~GeV).
All of these energy scales have been used to generate neutrino masses.

First note that one can generate Eq.~(\ref{5-dim}) without invoking
new mass scales above the weak scale, {\it i.e.}\/ using $M=M_W$ in
such a way that $\lambda^{ij}$ are very small.  One example is SUSY
with R-parity violation~\cite{Hirsch:2000ef}, where $\lambda^{ij}$ are
given as a product of very suppressed couplings.  Another possibility
is to make use of the GUT scale and $1/M_{Pl}$ operators to induce
right-handed neutrino masses of the order $M_{GUT}^2/M_{Pl}$, such that the
heaviest neutrino mass is of order 
$\frac{m_{top}^2M_{Pl}}{M_{GUT}^2}\simeq 0.1$~eV \cite{BPW}.

In the case of SUSY, it has recently been pointed out that one can
generate small Majorana or Dirac neutrino masses from SUSY breaking 
\cite{susynus}. In these models, the neutrino masses are small for the 
same reason that the SUSY-preserving $\mu$-term is much smaller than the 
unknown high energy scale (say $M_{Pl}$).

Finally, theories with large extra dimensions can generate very small
Dirac neutrino masses (Eq.~(\ref{yukawa})) by assuming that the SM
singlet fermions live in the extra dimensions and therefore their
couplings in the SM ``brane'' are volume suppressed~\cite{extra}.
This means that neutrino masses are small for the same reason that the
gravitational force is weak. Such theories, however, typically
require that $L$ be conserved, unless it is broken in distant branes
or by a totally different mechanism, such as R-parity
violation~\cite{4nubrane}.  This is required in order to forbid terms
similar to Eq.~(\ref{1/M_Pl}) but with $M=O(1)$~TeV, which would lead
to unacceptably large neutrino masses.

It is clear that small neutrino masses are ``expected'' on theoretical
grounds. However, the rich variety of theoretical options available
illustrates that neither their scale nor the magnitude of neutrino
mixing angles can be firmly predicted from first
principles~\cite{threview}.  Our goal here is to focus on what appears
to be the most economical way to solve the neutrino puzzles through
neutrino-mass-induced oscillations, ignoring indirect beyond the SM
hints. The motivation for this is two-fold: first, neutrino masses are
the only direct experimental hint for beyond the SM physics and
second, it is important to establish how contrived the new physics
has to be in order to explain the current solar and atmospheric
neutrino data.

\setcounter{footnote}{0}
\setcounter{equation}{0}
\section{The Minimalistic Model}

In order to minimize the amount of new physics invoked, we will take
advantage of the already present quantum gravity contribution to
neutrino masses, which na\"{\i}vely implies that all three neutrinos
weigh around $10^{-5}$~eV. Though not the favored solution, this is
enough to account for the current solar neutrino
data~\cite{latestglobalanalysis}. We need, however, to generate at
least one neutrino mass of order $10^{-1}$~eV in order to solve the
atmospheric neutrino puzzle. The most economical way of doing this is
to add {\sl one} right-handed neutrino, at the appropriate mass scale
$M_N$~\cite{Schechter:1980bn}.\footnote{The fact that a single
  right-handed neutrino can solve the neutrino puzzles has been explored
  in the context of SUSY theories \cite{SUSYsingle}.  Since these models 
  also require supersymmetry, they are far from ``minimalistic.''}
This point was also emphasised in \cite{King}, where a single right-handed
neutrino is used to explain the atmospheric data, while the solar data was
explained by GUT scale effects.

Why one right-handed neutrino, and not three? First of all, in order
to stick to the minimalistic approach, the smallest number of
right-handed neutrinos is to be added. Second, and most important,
there is, {\it a priori,} no correct number of right-handed neutrinos
since, as gauge singlets, they do not contribute to SM gauge
anomalies~\cite{Schechter:1980gr}. It is important to notice that these SM
singlet right-handed fermions are very heavy and decouple from the low
energy theory. Their only effect in the laboratory is indirect,
through the generation of a Majorana neutrino mass, even though, as will be 
discussed in the next section, they may also play some role in the generation 
of the baryon number of the universe.

The SM Lagrangian with $1/M_{Pl}$ operators and one right-handed
neutrino contains
\begin{equation}
\frac{\lambda^{ij}}{M_{Pl}}(L^iH)(L^jH) + y^i L^i N H + \frac{1}{2}M_N N N
+ h.~c.~.
\end{equation}
Integrating out the right handed fermion, the neutrino mass matrix is
(after electroweak symmetry breaking)
\begin{equation}
m_{\nu}^{ij}=\left(\frac{\lambda^{ij}}{M_{Pl}}+\frac{y^iy^j}{M_N}\right)v^2
\equiv M_1 + M_3.
\label{mass_min}
\end{equation}
The parameters $\frac{y^iy^j}{M_N}$ and $\lambda^{ij}$, which are of
order one, should be determined from the atmospheric and solar data.
In order to simplify things, we rewrite $\frac{y^iy^j}{M_N}v^2$ as
$m_3U_{i3}U_{j3}$, where $U_{e3}\equiv\sin\xi$,
$U_{\mu3}\equiv\cos\xi\sin\theta$, and
$U_{\tau3}\equiv\cos\xi\cos\theta$, such that $\sum_i|U_{i3}|^2=1$.
The reason for this notation will become clear shortly.

Neglecting the $1/M_{Pl}$ term, the neutrino mass matrix is
\begin{equation}
\mu_{\nu}=\left(\matrix{U_{e3}^2 & U_{e3}U_{\mu3} & U_{e3}U_{\tau3} \cr
 U_{\mu3}U_{e3} & U_{\mu3}^2 & U_{\mu3}U_{\tau3} \cr 
U_{\tau3}U_{e3} & U_{\tau3}U_{\mu3} & U_{\tau3}^2} \right)m_3.
\end{equation}    
This matrix has a projective
structure~\cite{ProjectiveMassMatrix,Hirsch:2000ef}, giving rise to
one nonzero eigenvalue $m_3$ and two zero eigenvalues.  However, since
lepton number has been broken, there is no reason for any of the neutrino
masses to be zero.  Indeed, it is easy to see that
nonzero masses to all neutrinos arise at the two-loop level.
However, these are significantly smaller than the contribution from
the $1/M_{Pl}$ operator we are considering here \cite{2loops}.

The eigenvector corresponding to the nonzero eigenvalue $m_3$ is
$\vec{v}_3\equiv (U_{e3},U_{\mu3},U_{\tau3})^{\top}$ in the flavour
basis which, because of unitarity, is determined by only two mixing
angles~\cite{Schechter:1980bn}, which we refer to as $\xi$ and
$\theta$. This nonzero eigenvalue is the one that will be used to
solve the atmospheric neutrino puzzle. If we choose $m_3=0.05$~eV,
$\sin\xi\simeq0$ and $\cos^2\theta\simeq\sin^2\theta\simeq 0.5$, both
the atmospheric neutrino data and the bounds from reactor neutrino
experiments~\cite{chooz} are easily satisfied.

We now proceed to show that the addition of the $1/M_{Pl}$ term easily
yields the just-so solution to the solar neutrino puzzle. For
illustrative purposes, we initially assume that the $1/M_{Pl}$ term is
``democratic,'' {\it i.e.}
\begin{equation}
m_{\nu}=\left(\matrix{U_{e3}^2 m_3+m & U_{e3}U_{\mu3} m_3+m & U_{e3}U_{\tau3} m_3+m \cr
 U_{\mu3}U_{e3} m_3+m & U_{\mu3}^2 m_3+m & U_{\mu3}U_{\tau3} m_3+m \cr 
U_{\tau3}U_{e3} m_3+m & U_{\tau3}U_{\mu3} m_3+m & U_{\tau3}^2 m_3+m} \right),
\label{demo}
\end{equation} 
where $m=\frac{\lambda v^2}{M_{Pl}}=\lambda \times 10^{-5}$~eV, and
$\lambda$ is some order one coefficient. It is easy to see that
Eq.~(\ref{demo}) has one zero eigenvalue. This is due to the
democratic nature of the $1/M_{Pl}$ operator. Note that one can
rewrite Eq.~(\ref{demo}) as
\begin{equation}
\label{2nuR}
m_{\nu}=\left(\matrix{U_{e3} & \frac{1}{\sqrt{3}} \cr U_{\mu3} & \frac{1}{\sqrt{3}} \cr
 U_{\tau3} & \frac{1}{\sqrt{3}} }\right) \left(\matrix{m_3 & 0 \cr  0 & 3m} \right)
\left(\matrix{U_{e3} & U_{\mu3} & U_{\tau3} \cr
 \frac{1}{\sqrt{3}} & \frac{1}{\sqrt{3}} & \frac{1}{\sqrt{3}} } \right),
\end{equation} 
in such a way that it mimics the mass matrix one obtains in models
with two right-handed neutrinos. Furthermore, in the case $m_3=0$, the
mass matrix has only one massive eigenvalue, $m'\equiv3m$ with
corresponding eigenvector $\vec{v}_1\equiv
(1/\sqrt{3},1/\sqrt{3},1/\sqrt{3})^{\top}$.

The neutrino mass matrix given by Eq.~\ref{demo} is simple to
diagonalize since, as mentioned before, it has a zero eigenvalue and
its corresponding eigenvector is trivial to compute. Note that
$m_{\nu}=M_3+M_1$, such that $M_3\vec{v_3}=m_3\vec{v_3}$ and
$M_1\vec{v_1}=m'\vec{v_1}$. Furthermore, $M_i\vec{v}^{\perp}=0$, where
${v}^{\perp}$ is any vector orthogonal to $\vec{v}_i$ ($i=1$ or 3).
Therefore, the massless eigenvector of $m_{\nu}$ is $\vec{V}_0\propto
\vec{v}_1\times\vec{v}_3$. The remaining eigenvectors are going to be
linear combinations of $\vec{v}_1$ and $\vec{v}_3$. It remains,
therefore, to solve
\begin{equation}
m_{\nu}\left(A \vec{v}_{1} + B \vec{v}_3\right)=(M_1+M_3)\left(A \vec{v}_{1} + B 
\vec{v}_3\right)=\Lambda \left(A \vec{v}_{1} + B \vec{v}_3\right),
\label{ABL}
\end{equation}   
for $\Lambda$, $A$, and $B$ (of course, $A$ and $B$ are related by requiring the 
eigenvectors to be normalised). Defining $\vec{v}_1\cdot\vec{v}_3\equiv\cos\zeta$,
the exact solution to Eq.~(\ref{ABL}) is
\begin{eqnarray}
\left(\frac{A}{B}\right)^{\pm} &=& \frac{\Lambda^{\pm}-m_3}{m_3\cos\zeta}, \\
\Lambda^{\pm} &=& \frac{m'+m_3}{2}\pm\frac{m'+m_3}{2}\left[1-\frac{4m'm_3}{(m'+m_3)^2}
\sin^2\zeta
\right]^{\frac{1}{2}}.
\end{eqnarray}
One can see that to first order in $m'/m_3$ (the limit of interest
here) the mass eigenvalues are, $\Lambda^+=m_3+m'\cos^2\zeta$,
$\Lambda^-=m'\sin^2\zeta$, and $\Lambda^0=0$, with corresponding
eigenvectors $\vec{V}^+\propto (\vec{v}_3 + \frac{m'}{m_3}\vec{v}_1)$,
$\vec{V}^-\propto (\vec{v}_1 - \cos\zeta\vec{v}_3)$ and
$\vec{V}^0\propto (\vec{v}_1\times\vec{v}_3)$.

Finally, we discuss what conditions must be met in order to solve the
neutrino puzzles.  First, the atmospheric data and the reactor
constraints are, as expected, solved by choosing $\vec{v}_3$ and $m_3$
as in the previous case where the effect of the $1/M_{Pl}$ term was
neglected ($m' = 0$).  The value of $\Delta m^2$ required for
the ``just-so'' solution to the solar neutrino puzzle is also easily
obtained for an order one $\lambda$: note that
$\sin^2\zeta=1-(U_{e3}+U_{\mu3}+U_{\tau3})^2/3$, which only vanishes
if all $U_{i3}=1/\sqrt{3}$, which is not allowed by the atmospheric
data~\cite{latestglobalanalysis}. In the extreme case of
$\sin^2\xi=0.1$ and $\sin^2\theta=0.5$, one has $\sin^2\zeta\simeq
0.1$, and $\Lambda^-\simeq0.3\lambda\times 10^{-5}$~eV, which is
$O(10^{-5})$ for an $O(1)$ value of $\lambda$.

A more serious concern is to guarantee large mixing in the solar
sector.  This can be read of from $U_{e1}\equiv
\cos{\xi}\cos{\omega}$, where $\omega$ is the ``solar'' mixing angle.
Ordering the mass eigenstates in the order of increasing mass,
$U_{e1}=(\vec{V}_0)_1=(U_{\mu3}-U_{\tau3})/(\sqrt{3}\sin{\zeta})$.
This vanishes if $U_{\mu3}=U_{\tau3}$, which implies $\cos\omega=0$,
which is unacceptable. The situation is easily remedied, however, by
requiring $U_{\mu3}$ and $U_{\tau3}$ to have opposite signs.

\begin{figure}
    \centerline{
    \psfig{file=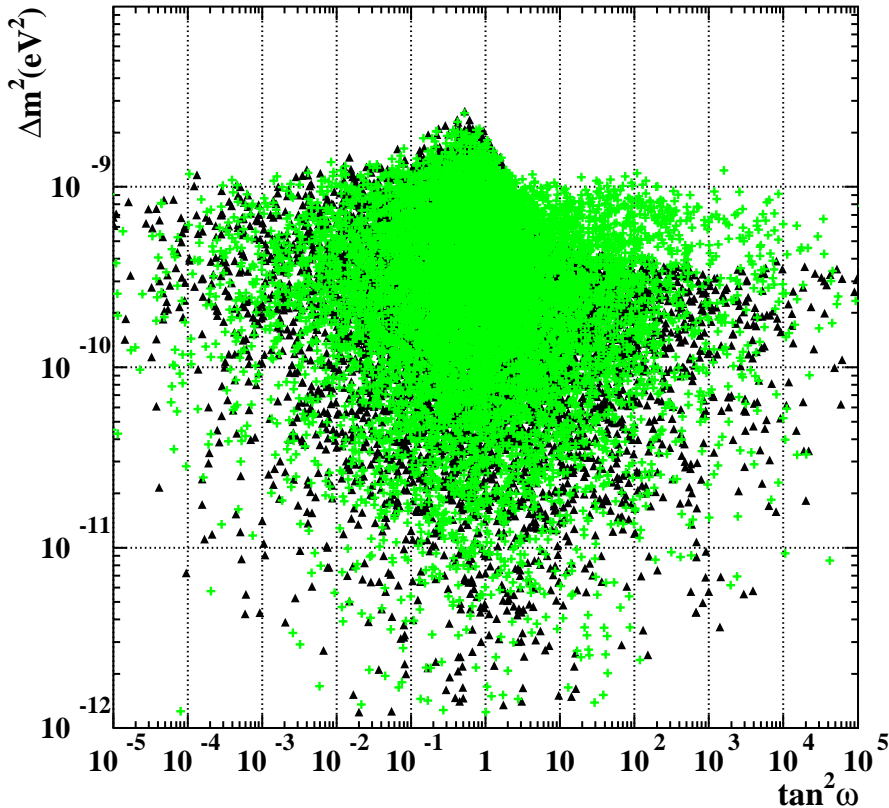,width=0.75\columnwidth}
        }
    \caption{10000 values of $\Delta m^2_{\odot}$ and $\tan^2\omega$ obtained by 
randomly generated elements for the $M_1$ matrix (see text), for $\sin^2\theta=0.5$, 
$\sin^2\xi=0$ (dark triangles) or $\sin^2\theta=0.5$, 
$\sin^2\xi=0.1$ (light crosses), and $m_3=0.05$~eV.}
    \label{dm2_t2o}
\end{figure}

Though there are (loose) arguments in favour of a democratic $1/M_{Pl}$
contribution to the neutrino Majorana mass matrix, there is no deep
reason for it. The agreement between the neutrino puzzles and the
model proposed here does not depend on democratic dimension-5
operators, but is a rather generic feature of random $1/M_{Pl}$
coefficients.  In order to check whether this is really the case, we
performed a numerical diagonalization of Eq.~(\ref{mass_min}) with
random $\lambda^{ij}$ matrices and choosing $y^i$ such that the the atmospheric
neutrino data is satisfied. Fig.~\ref{dm2_t2o} depicts
the values obtained for $\Delta m^2_{\odot}$ and $\tan^2\omega$ 
(where $\omega$ is the ``solar'' angle) for 10000 randomly generated $\lambda^{ij}$
(with elements between -2 and 2), for $\sin^2\theta=0.5$, 
$m_3=0.05$~eV and $\sin^2\xi=0$ 
(dark triangles) or $\sin^2\xi=0.1$ (light crosses). It should be noted that
close to maximal mixing ($\tan^2\omega\sim 1$) is preferred, and
that the values of $\Delta m^2_{\odot}$ obtained agree quite well with the small
$\Delta m^2_{\odot}$ range obtained in \cite{latestglobalanalysis}. 
The situation remains the same if extra complex phases are included at random.

\begin{figure}
    \centerline{
    \psfig{file=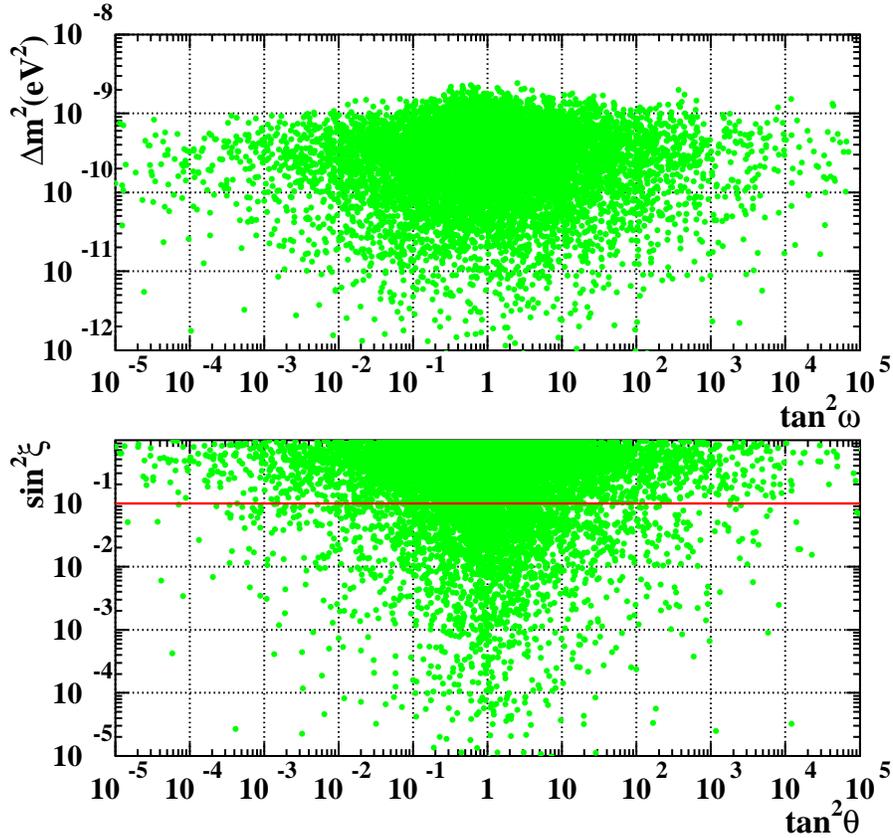,width=0.75\columnwidth}
        }
    \caption{10000 values of $\Delta m^2_{\odot}$ and $\tan^2\omega$ (top), and 
$\sin^2\xi$ $\tan^2\theta$ (bottom) obtained by 
randomly generated elements for the $M_1$ matrix and $U_{e3}$, $U_{\mu3}$, for
$m_3=0.05$~eV (see text). The solid line indicates $\sin^2\xi=0.1$, which is 
a good approximation for the upper bound on $|U_{e3}|^2$ from reactor, solar and
atmospheric neutrino data \cite{latestglobalanalysis}. }
    \label{all_four}
\end{figure}

One possible criticism to the minimalistic model is the fact that, even
though the solar data is naturally satisfied for random ``$1/M_{Pl}$''
operators, we are required to not only choose the mass scale for the
right-handed neutrino but also the relative values of the Yukawa couplings in
order to satisfy the atmospheric neutrino data.  This is not correct,
and we proceed to argue that randomly generated $U_{i3}$ easily
satisfy not only the atmospheric neutrino data but, to some extent,
also the reactor bounds~\cite{random}.  Fig.~\ref{all_four} depicts
10000 values for the two ``solar'' neutrino oscillation parameters
($\Delta m^2_{\odot}$ and $\tan^2\omega$, as before) and the
parameters $\sin^2\xi$ and $\tan^2\theta$, where $\theta$ and $\xi$
are the ``atmospheric'' and ``reactor'' angles, respectively, 
when not only $\lambda^{ij}$ but also
$U_{e3}$ and $U_{\mu3}$ and the sign of $U_{\tau3}$ are generated at
random (the absolute value of $U_{\tau3}$ is determined by $\sum_i
|U_{i3}|^2=1$). The only input to the mass matrix is the value of
$m_3=0.05$~eV, which is chosen such that the atmospheric data is
satisfied. The scatter plots in Fig.~\ref{all_four} can be directly
compared with the latest global three-neutrino analysis of the
neutrino oscillation data \cite{latestglobalanalysis}. Note that, even
though large values for $\sin^2\xi$ are preferred, a significant fraction
of the points falls within the allowed region $\sin^2\xi\lesssim 0.1$.

We summarize this section by restating that, as expected, the
minimalistic neutrino mass model can easily account for the solar and
atmospheric neutrino puzzles.  The ``light'' neutrino masses and the
``solar'' mixing angle $\omega$ come from $v^2/M_{Pl}$ and order one
coefficients, which can be picked almost at random. The
``atmospheric'' angle $\theta$ and the ``reactor'' angle $\xi$ are
related to the Yukawa couplings of the right and left-handed
neutrinos. We have shown that, if chosen randomly, these can easily
account for the data.  The only required input to the model is the
value of $m_3= (yv)^2/M_N$, which has to be chosen such that the
atmospheric neutrino data is satisfied.

\setcounter{footnote}{0}
\setcounter{equation}{0}
\section{Leptogenesis}

One of the exciting feature of models with heavy Majorana fermions is
the fact that they potentially contain all the ingredients required to
explain the small but nonzero baryon-to-entropy ratio of the universe:
$n_B/s\sim 10^{-10}$. This is accomplished by the mechanism of
leptogenesis \cite{Leptogenesis}. In a nutshell \cite{Leptoreview},
because the heavy ``right-handed neutrinos'' are Majorana particles,
their decays violate lepton number. If these decays also violate CP
and happen out of thermal equilibrium, a net lepton number for the
universe is generated. Later, through sphaleron effects
\cite{sphalerons}, the net lepton number is partially transformed into
a net baryon number, as long as the lepton number is produced before
the electroweak phase transition ($T\simeq 10^3$~GeV). If this is the
case, the net baryon-to-entropy ratio has been estimated
\cite{lintob} to be
\begin{equation}
\frac{n_B}{s} = \left(\frac{24+4N_H}{66+13N_H}\right)\frac{n_{B-L}}{s},
\end{equation}
where $N_H$ is the number of Higgs doublet and $n_{B-L}$ is the amount
of baryons minus leptons. In the SM, $N_H=1$ and
$n_B/s=(28/79)n_{B-L}/s$, or $n_B/s=-(28/79)n_L/s$, assuming that
initially only a net lepton number is generated.  In this section, we
will discuss whether our minimalistic model is capable of generating
the observed baryon-to-entropy ratio.

\begin{figure}
    \centerline{
    \psfig{file=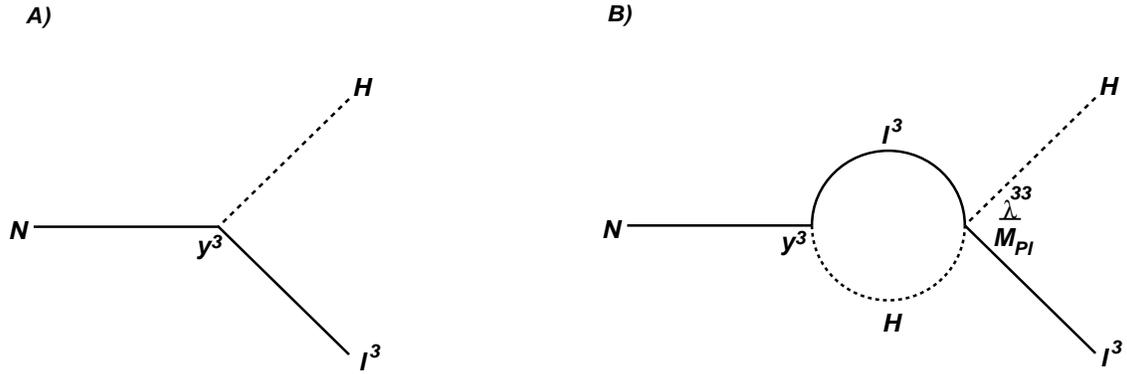,width=1\columnwidth}
        }
    \caption{Feynman diagrams for the decay $N \to l^3H$: (A) 
      dominant tree-level contribution and (B) one-loop contribution
      containing the $M_{Pl}$-suppressed operator.}
    \label{decay}
\end{figure}

In order to estimate ``how much'' lepton number is produced during
right-handed neutrino decays, we compute \cite{Leptoreview}
\begin{equation}
\epsilon=\frac{\Gamma(N\rightarrow l^3H)-\Gamma(N\rightarrow \bar{l^3}H^*)}
{\Gamma(N\rightarrow l^3H)+\Gamma(N\rightarrow \bar{l^3}H^*)}.
\end{equation}
A nonzero value for $\epsilon$ arises from the interference between
the tree-level and the absorptive part of the one-loop diagram in
Fig.~\ref{decay}, and is given by
\begin{equation}
\epsilon=\frac{\Im(y_3^2\lambda_{33}^*)}{|y_3|^28\pi}
\frac{M_N}{M_{Pl}}
\simeq \frac{mM_N}{8\pi v^2}\sin\delta,
\end{equation}
where $y_3$ is defined as the Yukawa coupling between the SM
singlet-fermion ($N$), and the appropriate linear combination of SM
lepton doublets ({\it i.e.} $\mathcal{L}\supset y_3 L_3 NH$),
$\lambda_{33}$ is the coefficient of the $1/M_{Pl}$ operator when the
lepton doublets are expressed in a basis which contains $L_3$, $m =
O(10^{-5})$~eV is the light neutrino mass, $v$ is the SM Higgs vev,
$M_N$ is the Majorana mass of the right-handed state, and $\delta$ is
the phase of $y_3^2\lambda_{33}^*$.  Numerically
\begin{equation}
\epsilon=10^{-7}\left(\frac{m}{10^{-5}~\rm eV}\right)\left(\frac{M_N}{10^{13}~\rm GeV}
\right)\left(\frac{\sin\delta}{1}\right).
\label{number_e}
\end{equation} 

Note that a nonzero value for $\epsilon$ is obtained, even though
there is only one right-handed neutrino species. It is clear that,
within the renormalizable Lagrangian, it is impossible to generate a
net lepton number with just one right-handed neutrino species, due to
the absence of CP-violation. However, the presence of the dimension-5
operator solves this problem.
This is not at all surprising since the situation here is analogous to
having two (or more) right-handed neutrinos with hierarchical Majorana
masses (see. {\it e.g.,}\/ Eq.~(\ref{2nuR})). In this case, it is well known that a
net lepton number can be generated via the decay of the lightest
right-handed neutrino state.

Is Eq.~(\ref{number_e}) enough to generate the observed
baryon-to-entropy ratio?  Unfortunately, the answer depends on the
history of the very early universe. The simplest hypothesis, perhaps,
is to assume that below some very high temperature the universe is
described by a thermal bath of all SM particles, including the heavy
right-handed neutrino. In this case, the question which must be
addressed is whether the right-handed neutrino falls out of thermal
equilibrium before decaying. Such a condition is (roughly) satisfied
if the right-handed neutrino decay width is smaller than the expansion
rate of the universe at temperatures $T\sim M_N$. From
Fig.~\ref{decay}(A), the decay width of the right-handed neutrino is
\begin{equation}
\Gamma=\frac{y_3^2}{8\pi}M_N\simeq\frac{m_3 M_N^2}{8\pi v^2}
\end{equation}
where $m_3 \simeq (y_3v)^2/M_N$ is the heaviest neutrino mass, chosen
to account for the atmospheric mass scale. In a radiation dominated
universe ($H \simeq 1.7g_*^{1/2}T^2/M_{Pl}$), the out-of-equilibrium
constraint implies
\begin{equation}
m_3\lesssim 10^{-3}~{\rm eV}.
\end{equation}
which is about two orders of magnitude below the required $m_3 \sim
0.05$~eV, implying that the right-handed neutrinos do not decay
sufficiently out of thermal equilibrium and that the leptogenesis
mechanism cannot generate the observed baryon-to-entropy ratio. More
detailed estimates reach the same conclusion. For example, the authors
of Ref.~\cite{reheat} estimate that values of $\epsilon\sim 10^{-5}$
are required for values of $m_3\lesssim 10^{-2}$~eV, which cannot be
satisfied by our minimalistic model.

%
One simple way of circumventing this is to assume a different history
for the early universe. In particular, if the right-handed neutrinos
are never in equilibrium with the thermal bath of light states, the
above constraint does not apply. Estimates of the number density of
right-handed neutrinos at the time of their decay and the resulting
baryon-to-entropy ratio are very model dependent, and require that we go
beyond the minimalistic model (indeed, we have nothing to say about inflation, 
for example).
Nonetheless, it is worth mentioning that, for example, in a number of
inflationary models, right-handed neutrinos may be produced at the
time of reheating even if the reheating temperature $T_{RH}$ is (much)
lower than its mass (see \cite{reheat} and references therein for
details). In this case, they are almost always (way) out of
equilibrium when they decay, and the estimates for the generated
baryon-to-entropy ratio can be drastically different.  Two
possibilities were analyzed in detail in \cite{reheat}. First, the
right-handed neutrinos may be coupled to the inflaton field and
therefore are still produced at the time of reheating even if
$T_{RH}<M_N$, as long as they are lighter than the inflaton field
itself. Second, the right-handed neutrinos may be produced during
preheating \cite{preheating}, even with relatively low reheating
temperatures ($M_N\gg T_{RH}$) due to the non-perturbative decay of
large inflaton oscillations during reheating. This mechanism is
particularly efficient for fermions \cite{reheat}.
In both cases it seems plausible that our minimalistic neutrino mass
model can generate the observed baryon-to-entropy ratio. According
to estimates of \cite{reheat}, this will happen as long as $M_N\gtrsim
10^{13}$~GeV and the reheating temperature is ``large'' ($T_{RH}
\gtrsim 10^{10}$~GeV).

\setcounter{footnote}{0}
\setcounter{equation}{0}
\section{Discussions and Conclusions}

The SuperKamiokande atmospheric neutrino data strongly suggest that
effects beyond the standard model of particle physics are present in
the leptonic sector. The simplest extension of the SM which explains the
atmospheric neutrino data is to assume that neutrinos are massive, and
that they mix.  In light of this strong experimental evidence,
it is natural to assume that the long standing solar neutrino puzzle
is also solved by neutrino oscillations. 

In this paper, we have emphasized that nonrenormalizable $1/M_{Pl}$
operators provide a solution to solar neutrino puzzle without the
addition of any new physics. On the other hand, the atmospheric
neutrino puzzle {\sl requires} the existence of new physics at
intermediate energy scales. The most ``economical'' model requires one
new field at this new physics scale, which can be anywhere between
$10^9$~GeV and $10^{14}$~GeV for typical Yukawa coupling values. It is
important to note that, {\sl as far as solving the neutrino puzzles is
  concerned}, this simple model is just as successful as other more
complex models. A potential drawback of the model is that mixing angles are not
predicted, so that the required relative magnitudes of the Yukawa
coupling has to be put in by hand in order to satisfy the atmospheric
neutrino data. In particular the right-handed neutrino needs to couple to
$\nu_{\mu}$ and $\nu_{\tau}$ with equal strength, while the coupling
to $\nu_e$ has to be somewhat suppressed.
However, as we have shown (Fig.~\ref{all_four}), 
the present experimental constraints on 
$|U_{e3}|\lsim 0.3$ can still be accommodated without 
difficulty by assuming that the Yukawa couplings are picked completely
at random, in the spirit of neutrino mass ``anarchy'' \cite{random}. Within
this perspective, all mixing angles easily arise from random Yukawa couplings and
random coefficients for the $1/M_{Pl}$ operator, and the only input to the model
is the value of $m_3$. This situation may well change in the near future if $U_{e3}$ 
is constrained to be very small. 

The idea that the solar neutrino puzzle may be evidence for ``quantum
gravity'' effects will be severely tested in the near future.  The
``just-so'' solution ($\Delta m^2_{\odot} \lesssim 10^{-10}$~eV$^2$) is
already somewhat disfavoured by the solar neutrino
data~\cite{latestglobalanalysis}, while values of $\Delta
m^2_{\odot} \gtrsim 5 \times 10^{-10}$~eV$^2$ are still acceptable
\cite{latestglobalanalysis} in the recently revealed ``quasivacuum''
region \cite{QV}. However, the Borexino experiment, which is to start
data taking next year, will thoroughly explore $\Delta m^2_{\odot} \lesssim
5\times 10^{-9}$~eV$^2$ and large mixing in a standard solar model independent way
\cite{seasonal}. It is safe to say that, after a few years, these values
of $\Delta m^2_{\odot}$ will be either confirmed or definitively ruled out.
Furthermore, the minimalistic model also predicts
hierarchical neutrino masses, such that the sign of the
``atmospheric'' mass-squared difference $\Delta m^2_{\rm atm}$ is
positive. This possibility will be definitively tested at a future neutrino factory 
\cite{nufact}.

\section*{Acknowledgements}

We would like to thank Martin Hirsch, Steve King, Magda Lola, and
Apostolos Pilaftsis for useful and enlightening conversations, while
AdG acknowledges the hospitality of IFIC during his stay in
Val\`encia, where this work was initiated.  This work was supported by
DGICYT under grant PB98-0693 and by the TMR network grants
ERBFMRXCT960090 and HPRN-CT-2000-00148 of the European Union.


\begin{thebibliography}{99}

\bibitem{SuperK-atmos} 
  SuperKamiokande Collaboration, 
  Y.~Fukuda {\it et al.}, {\sl Phys. Lett.}\/ {\bf B433}, 9 (1998);
  {\sl Phys. Lett.}\/ {\bf B436}, 33 (1998);
%
  {\sl Phys. Lett.}\/ {\bf B467}, 185 (1999);  
  {\sl Phys. Rev. Lett.}\/ {\bf 82}, 2644 (1999). 
%
  H.~Sobel, talk at  XIX International Conference on Neutrino Physics 
  and Astrophysics, Sudbury, Canada, June 2000   
  ({\it http://{\-}nu2000.{\-}sno.{\-}laurentian.{\-}ca}); 
  T.~Toshito, talk at the
  XXXth International Conference on High Energy Physics, July 27 -
  August 2, 2000 (ICHEP 2000) Osaka, Japan
  ({\it http://{\-}www.{\-}ichep2000.{\-}rl.{\-}ac.{\-}uk}). 
 
\bibitem{solar} 
%
 Y.~Suzuki, talk  at XIX International Conference on Neutrino Physics 
 and Astrophysics, Sudbury, Canada, June 2000   
  ({\it http://{\-}nu2000.{\-}sno.{\-}laurentian.{\-}ca}); 
  T.~Takeuchi, talk at the
  XXXth International Conference on High Energy Physics, July 27 -
  August 2, 2000 (ICHEP 2000) Osaka, Japan
  ({\it http://{\-}www.{\-}ichep2000.{\-}rl.{\-}ac.{\-}uk}). 
%
 Homestake Collaboration, B.T.~Cleveland {\it et al.},
 {\sl Astrophys. J.}\/ {\bf 496}, 505 (1998); 
 R.~Davis, {\sl Prog. Part. Nucl. Phys.}\/ {\bf 32}, 13 (1994);
 K.~Lande, talk at XIX International Conference on Neutrino Physics 
  and Astrophysics, Sudbury, Canada, June 2000 
  ({\it http://{\-}nu2000.{\-}sno.{\-}laurentian.{\-}ca}); 
%
SAGE Collaboration, J.N. Abdurashitov {\it et al.},
 {\sl Phys. Rev.}\/ {\bf C60}, 055801 (1999); V.~Gavrin, talk at 
 XIX International Conference on Neutrino Physics and Astrophysics,
 Sudbury, Canada, June 2000 
 ({\it http://{\-}nu2000.{\-}sno.{\-}laurentian.{\-}ca}). 
%
GALLEX Collaboration, W.~Hampel {\it et al.},
 {\sl Phys. Lett.}\/ {\bf B447}, 127 (1999).
%
E.~Belloti, talk at 
 XIX International Conference on Neutrino Physics and Astrophysics,
 Sudbury, Canada, June 2000 
 ({\it http://{\-}nu2000.{\-}sno.{\-}laurentian.{\-}ca}). 

\bibitem{atmos} 
%
  R.~Becker-Szendy {\it et al.}, {\sl Phys. Rev.}\/ {\bf D46}, 3720 (1992). 
%
  H.S.~Hirata {\it et al.}, {\sl Phys. Lett.}\/ {\bf B280}, 146 (1992); 
  Y.~Fukuda {\it et al.}, {\it ibid} {\bf B335}, 237 (1994).
%
W.W.M.~Allison {\it et al.,}\/ {\sl Phys. Lett.}\/ {\bf
    B449}, 137 (1999). A.~Mann, talk at XIX International Conference
  on Neutrino Physics and Astrophysics, Sudbury, Canada, June 2000
  {\it  http://nu2000.sno.laurentian.ca/T.Mann/index.html}
%
  2000, {\it http://www.ichep2000.rl.ac.uk/Program.html}. See also
  hep-ex/0001058, Proc.  of the {\sl Sixth International Workshop on
    Topics in Astroparticle and Underground Physics}, TAUP99, Paris
  September 1999.
%

\bibitem{solrsf}
  O.~Miranda, {\it et al.,}\/ hep-ph/0005259 ({\sl Nucl. Phys.}\/ 
  {\bf B} in press).

\bibitem{solexotic} S.~Bergmann, {\it et al.,}\/ 
  {\sl Phys. Rev.}\/ {\bf D62}, 073001 (2000).

\bibitem{atmexotic}
N.~Fornengo, M.C.~Gonzalez-Garcia, and J.W.F.~Valle, {\sl JHEP}\/ {\bf
  0007}, 006 (2000); M.C.~Gonzalez-Garcia {\it et al.}, {\sl Phys. Rev. Lett.}\/
{\bf 82}, 3202 (1999) and references therein.

\bibitem{oscreview} 
%
  For a recent review on the phenomenology of neutrino oscillations see,{\it e.g.,}\/
  S.M.~ Bilenky, C. Giunti, and W. Grimus, {\sl Prog. Part. Nucl. Phys.}\/ 
  {\bf 43} (1999) 11.
  
\bibitem{Valle:1987gv}
%
  J.W.F.~Valle,
{\sl Phys. Lett.}\/ {\bf B199} (1987) 432.

\bibitem{latestglobalanalysis}
M.C.~Gonzalez-Garcia, {\it et al.,}\/
hep-ph/0009350 and references therein. 

\bibitem{PDG} D.E.~Groom {\it et al.,}\/ {\sl Eur. Phys. J.}\/ {\bf
    C15}, 1 (2000).

\bibitem{double-beta} L.~Baudis {\it et al.,}\/ {\sl Phys. Rev.
    Lett.}\/ {\bf 83}, 41 (1999).


\bibitem{LSND} LSND Collaboration, C. Athanassopoulos et al., 
{\sl Phys. Rev. Lett.}\/ {\bf 75} (1995) 2650; {\sl Phys. Rev. Lett.}\/ 
{\bf 77} (1996) 3082; {\sl Phys. Rev. Lett.}\/ {\bf 81} (1998) 1774.


\bibitem{4nu} J.~T.~Peltoniemi, D.~Tommasini, and J.~W.~F.~Valle,
  {\sl Phys. Lett.}\/ {\bf B298} (1993) 383; D.~O.~Caldwell and
  R.~N.~Mohapatra, {\sl Phys. Rev.}\/  {\bf D48} (1993) 329; J.~T.~Peltoniemi
  and J.~W.~F.~Valle, {\sl Nucl. Phys.}\/ {\bf B406} (1993) 409
  [hep-ph/9302316].
  
\bibitem{4nubrane} A.~Ioannisian and J.W.F.~Valle,
  hep-ph/9911349; M.~Hirsch and J.W.F.~Valle, hep-ph/0009066, to
  appear in {\sl Phys.  Lett.}\/ {\bf B}.

  
\bibitem{Schechter:1980gr} J.~Schechter and J.~W.~F.~Valle, {\sl Phys. 
  Rev.}\/ {\bf D22}, 2227 (1980).

\bibitem{threview}
%
  For reviews on the theory of neutrino masses see, {\it e.g.,}\/
 J.W.F.~Valle, ``Neutrino Masses: From Fantasy to Facts,'' 
  Springer Tracts in Modern Physics
  163 (2000) 35-68, [hep-ph/9906378] and {\sl Prog. Part. Nucl. Phys.}\/ 
  {\bf 26} (1991) 91.

\bibitem{pdecay} Y.~Hayato {\it et al.,}\/ {\sl Phys. Rev. Lett.}\/ {\bf 
83}, 1529 (1999).

\bibitem{BEG} R.~Barbieri, J.~Ellis, and M.K.~Gaillard, {\sl Phys. Lett.}\/
{\bf B90}, 249 (1980).

\bibitem{ABS} E.Kh.~Akhmedov, Z.G.~Berezhiani and G.~Senjanovic, {\sl Phys.
Rev. Lett.}\/ {\bf 69}, 3013 (1992).

\bibitem{seesaw} M.~Gell-Mann, P.~Ramond, R.~Slansky, in {\sl
Supergravity}, ed. P.~van~Niewenhuizen and D.~Freedman (North
Holland, 1979); T.~Yanagida, in {\sl KEK lectures}, ed.  O.~Sawada and
A.~Sugamoto (KEK, 1979); R.N.~Mohapatra and G.~Senjanovic,
{\sl Phys. Rev. Lett.}\/ {\bf 44} 912 (1980).

\bibitem{ADD}
I.~Antoniadis,
{\sl Phys. Lett.}\/ {\bf B246} (1990) 377;
N.~Arkani-Hamed, S.~Dimopoulos, and G.~Dvali, {\sl Phys. Let.}\/ {\bf B429}, 263
(1998);
I.~Antoniadis, {\it et al.,}\/ 
{\sl Phys. Lett.}\/ {\bf B436}, 257 (1998) [hep-ph/9804398];
N.~Arkani-Hamed, S.~Dimopoulos, and G.~Dvali, {\sl Phys. Rev.}\/ {\bf D59},086004
(1999).

\bibitem{Hirsch:2000ef}
M.~Hirsch {\it et al.,}\/
hep-ph/0004115 to appear in {\sl Phys. Rev.}\/ {\bf D}; J.C. Rom\~ao, {\it et
al.,}\/ {\sl Phys. Rev.}\/ {\bf D61} (2000) 071703 and references therein.

\bibitem{BPW} K.S.~Babu, J.C.~Pati, and F.~Wilczek, {\sl Nucl. Phys.}\/ 
{\bf B566}, 33 (2000) and references therein.  

\bibitem{susynus} 
%
N.~Arkani-Hamed, {\it et al.,}\/ hep-ph/0006312.

\bibitem{extra} 
K.R. Dienes, E. Dudas, and T. Gherghetta, {\sl Nucl. Phys.}\/ {\bf B557}, 25 (1999);
N.~Arkani-Hamed, {\it et al.}, hep-ph/9811448;
N.~Arkani-Hamed, S.~Dimopoulos, hep-ph/9811353. 

\bibitem{Schechter:1980bn} J.~Schechter and J.W.F.~Valle, {\sl Phys. Rev.}\/ 
  {\bf D21}, 309 (1980).

\bibitem{SUSYsingle} S.~Davidson and S.~King, {\sl Phys. Lett.}\/ {\bf B445},
191 (1998).

\bibitem{King} S.~King, {\sl Phys. Lett.}\/ {\bf B439}, 350 (1998).

\bibitem{ProjectiveMassMatrix} A.~Santamaria and J.W.F.~Valle, {\sl Phys. 
  Lett.}\/ {\bf B195}, 423 (1987).

\bibitem{2loops} K.S.~Babu and E.~Ma, {\sl Phys. Rev. Lett.}\/ {\bf 61}, 
674 (1988).

\bibitem{chooz} M. Apollonio {\it et al.,}\/ {\sl Phys. Lett.}\/ 
{\bf B466} 415 (1999) [hep-ex/9907037]; F. Boehm {\it et al.,}\/ 
hep-ex/9912050.

\bibitem{random} 
L.~Hall, H.~Murayama, and N.~Weiner, {\sl Phys. Rev. Lett.}\/ 
{\bf 84}, 2572 (2000); N.~Haba and H.~Murayama, hep-ph/0009174.  
 
\bibitem{Leptogenesis} M.~Fukugita and T.~Yanagida, {\sl Phys. Lett.}\/ 
{\bf B174}, 45 (1986). 

\bibitem{Leptoreview} For recent reviews, see A.~Pilaftsis, 
{\sl Int. J. Mod. Phys.}\/ {\bf A14}, 1811 (1999); W.~Buchm\"uller and 
M.~Pl\"umacher, {\sl Phys. Rep.}\/ {\bf 320}, 329 (1999); hep-ph/0007176.

\bibitem{sphalerons} V.A.~Kuzmin, V.A.~Rubakov, and M.E.~Shaposhnikov, {\sl 
Phys. Lett.}\/ {\bf B155}, 36 (1985).

\bibitem{lintob} S.Y.~Khlebnikov and M.E.~Shaposhnikov, {\sl Nucl. Phys.}\/
{\bf B308}, 885 (1988); 
%
%
J.A.~Harvey and M.S.~Turner, {\sl Phys. Rev.}\/ {\bf D42},
3344 (1990).

\bibitem{reheat} 
%
  G.F.~Giudice {\it et al.,}\/ {\sl JHEP}\/ {\bf 9908}, 014 (1999).

\bibitem{preheating} 
%
  L.A.~Kofman, A.D.~Linde, and A.A.~Starobinsky, {\sl Phys. Rev.
    Lett}\/ {\bf 73}, 3195 (1994).


\bibitem{QV} 
%
A.~Friedland, {\sl Phys. Rev. Lett.}\/ {\bf 85}, 936 (2000); 
A.~de~Gouv\^ea, A.~Friedland, and H.~Murayama, {\sl Phys. Lett.}\/
{\bf B490}, 125 (2000); 
E.~Lisi {\it et al.,}\/ hep-ph/0005261.

\bibitem{seasonal} 
%
  A.~de~Gouv\^ea, A.~Friedland, and H.~Murayama, {\sl Phys. Rev.}\/
  {\bf D60}, 093011 (1999).

\bibitem{nufact}
V.~Barger, {\it et al.,}\/
{\sl Phys. Lett.}\/  {\bf B485} (2000) 379
[hep-ph/0004208]; A.~De~Rujula, M.B.~Gavela, and P.~Hernandez,
{\sl Nucl. Phys.}\/ {\bf B547} (1999) 21 [hep-ph/9811390]; M.~Freund,
P.~Huber, and M.~Lindner,
{\sl Nucl. Phys.}\/  {\bf B585} (2000) 105
[hep-ph/0004085].

\end{thebibliography}
\end{document}